\documentclass[article,12pt]{elsarticle}
\usepackage{amssymb}
\usepackage{xcolor}
\usepackage[english]{babel}
\usepackage[T2A]{fontenc}
\usepackage[utf8]{inputenc}
\journal{PEPAN Letters}

\newcommand{\ba}{\begin{eqnarray}}
\newcommand{\ea}{\end{eqnarray}}
\newcommand{\beqs}{\begin{eqnarray}}
\newcommand{\eeqs}{\end{eqnarray}}

\begin{document}

\begin{frontmatter}

\title{ {\bf On matter and pressure distribution in nucleons}}

\author{Roberto Fiore$^{a,}$\footnote{e-mail: roberto.fiore@cs.infn.it},  L\'aszl\'o Jenkovszky$^{b,}$\footnote{e-mail: jenk@bitp.kiev.ua}, Maryna Oleksiienko$^{c,}$\footnote{e-mail: maryna.oleksiienko@gmail.com}}
\address{{$^a$\sl Dipartimento di Fisica, Universit\'a della Calabria and \par
Istituto Nazionale di Fisica Nucleare, Gruppo collegato di Cosenza,\par
Rende, I-87036, Italy\par  
$^b$\sl Bogolyubov Institute for Theoretical Physics (BITP), National Academy of Sciences of Ukraine, Kyiv, UA-03143, Ukraine\par
$^c$ Faculty of Physics, Taras Shevchenko National University of Kyiv, Kyiv, UA-03680, Ukraine}}

\begin{abstract} Matter and pressure distribution in hadrons are studied in a dual analytic model of generalized parton distributions with complex Regge trajectories. An original parametrization of the pressure distribution in the nucleon is proposed, ensuring its stability and compatible with the experimental data from the JLab accelerator.
   \end{abstract}


\end{frontmatter}

\newpage

\section{Introduction} \label{s1}

Recently  the gravitational form factors (GFFs) became an important tool to probe the internal structure of hadrons  providing important
 information on the hadron’s mass, spin and
their mechanical property. In experiments, the GFFs can be constrained by the generalized parton distributions (GPDs) measured in
the hard exclusive processes \cite{Mueller, Rad, Ji}. Interest in the subject has been triggered by the recent paper \cite{Burkert_Elouadrhiri_Girod} where the quark GFFs were related to the pressure distribution inside the proton.

\begin{figure}[h] 
	\centering
\includegraphics[width=1\textwidth]{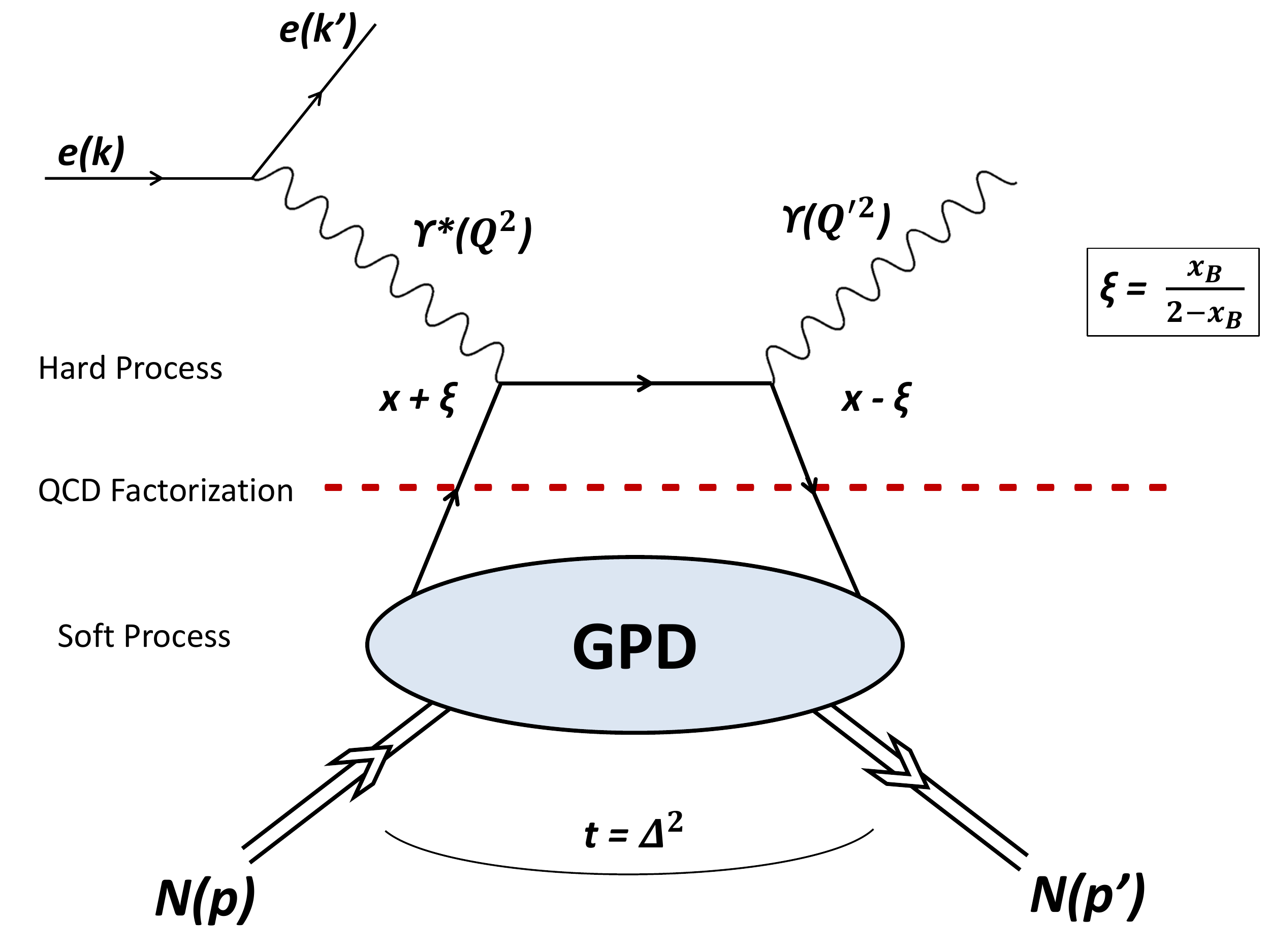}
	\caption{Diagram of deeply virtual Compton scattering (DVCS) with QCD factorization separating nonperturbative "soft"  and perturbative "hard" dynamics.} 
	\label{Fig:diagram}
\end{figure}

Generalized parton distributions (GPDs) \cite{Mueller, Rad, Ji, Jenk1, Jenk2} summarize our knowledge about the one-dimensional longitudinal parton momenta distributions with the transverse distribution of the nuclear matter, as seen in deeply virtual Compton scattering (DVCS), $\gamma^*(q) + N(p) \rightarrow \gamma(q^\prime) + N(p^\prime)$, Fig. \ref{Fig:diagram}. This figure contains also complete information and notation on kinematics used in the present paper (see, for instance, Sec. \ref{Sec:GPD}).     

The basic object in the formalism is the energy-momentum tensor (EMT), whose matrix elements according to quantum chromodynamics (QCD) can be parametrized as 

\begin{equation}
\langle p'\vert T_{\mu\nu}(0)\vert p\rangle=\overline{u}(p')[A(t)\frac{P_{\mu}P_{\nu}}{M_N}+J(t)\frac{i(P_{\mu}\sigma_{\nu\rho}+P_\nu\sigma_{\mu\rho})\Delta^\rho}{2M_N}+$$ $$+d(t)
\frac{\Delta_{\mu}\Delta_{\nu}-g_{\mu\nu}\Delta^2}{4M_N}]u(p),
\end{equation}
where $P=(p+p')/2$ and$\ \ \Delta=(p'-p)$ are related to the familiar Mandelstam variables $4P^2=s, \ \  \Delta^2=t$.  

GFFs are matrix elements of the EMT, where  
 $A(t), \ J(t)$ and $d(t)$  correspond respectively to the mass, angular momentum (spin) and internal pressure distributions. In Ref. \cite{Teryaev2} the elements of the EMT were related to moments of the GPDs, responsible of the momentum distribution in nuclei. 
 
As anticipated in Ref. \cite{Pagels} (see also  Refs. \cite{Weinberg, Landau, Chugreev}), the fundamental properties of particles, such as their masses or spin, may be interpreted as the particle response to the external gravitational field (modification of the space-time metric). 

Recently, the third term in the EMT, related to  the internal pressure, is in the focus of theorists and experimentalists (see below, Sec. \ref{Sec:Pressure} of the present paper). 

Our formalism is based on the analytic $S-$matrix theory properties, in particular its realization in the Regge-pole and dual theories. In our opinion, Regge trajectories contain much information on the dynamics of strong interactions. It is important that they are non-linear complex functions. We dedicate a  
special section introducing this object.

\section{Non-linear, complex Regge trajectories}\label{sec:unandu}
\begin{figure}[h] 
	\centering
\includegraphics[width=1\textwidth]{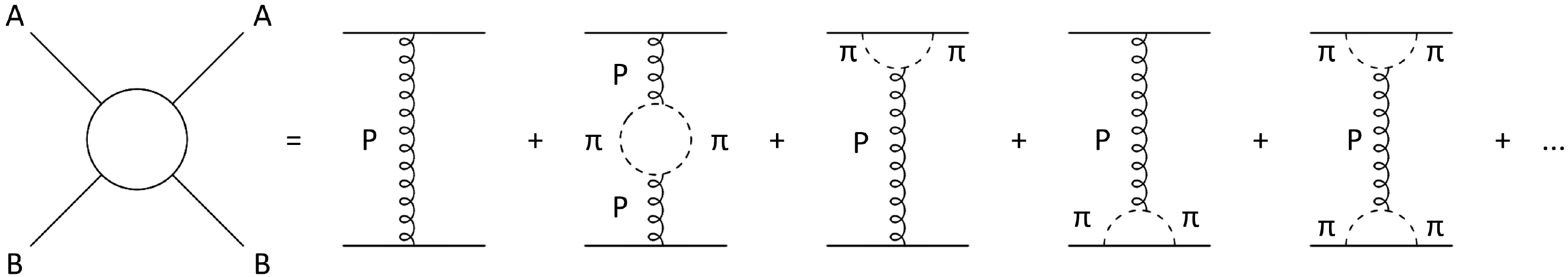}
	\caption{Regge pole (here, pomeron) exchange diagram with loops generated by the two-pion threshold, Eq. (\ref{hr}).} 
	\label{Fig:Diagram}
\end{figure}

Unitarity constrains the threshold behaviour of Regge trajectories. 
Fig. \ref {Fig:Diagram} shows a typical diagram corresponding to the exchange of the vacuum (Pomeranchuk, or pomeron) trajectory   
\begin{equation} \label{hr}
\Im m\alpha(t)_{t \rightarrow t_0}\sim(t-t_0)^{\Re e\alpha(t_0)+1/2}.
\end{equation}

Asymptotically, the trajectories  must obey the upper limit 
\begin{equation}\label{asympt}
\Bigg\vert\frac{\alpha(t)}{\sqrt{t}\ln{t}}\Bigg\vert_{t\rightarrow\infty}\leq {\rm const}.
\end{equation}

The simple trajectory 
 \citep{Trajectory}
\begin{equation}\label{3}
    \alpha(t) = \frac{1+\delta+\alpha_{1}t}{1+\alpha_{2}\big(\sqrt{t_{0}-t}-\sqrt{t_0})}
\end{equation}
satisfies the above constrains.
Here $t_0=4m_{\pi}^2$  or $t_0=9m_\pi^2$ 
for the pomeron or the odderon (odd-$C$ counterpart of the pomeron), respectively and
$\delta, \alpha_1, \alpha_2$ are adjustable parameters.

The threshold in the trajectory (and consequently in the amplitude) bears  the deep physical meaning by creating the nucleon atmosphere, as shown in Fig. \ref{Fig:3}.

For $|t| \texttt{>>} t_0$ it is$\ \  |\alpha(t)|\rightarrow({\alpha_1}/{\alpha_2})\sqrt{|t|}$. The intercept and the forward slope of the trajectory are respectively $\alpha(0)=1+\delta$ and  
\begin{equation}\label{Slope}
\alpha'(0)=\alpha_1+\alpha_2\frac{1+\delta}{2\sqrt{t_0}}. 
\end{equation}    

\section{Matter distribution}\label{Sec:Matter}
Unlike the distribution of the electric charge, evident from electron-nucleon scattering \cite{Vorobyev, Chugreev}, the distribution of the nuclear matter is not seen directly. Attempts to infer it by means of GPDs, e.g. as in Ref. \cite{Selugin}, resulted to a rather trivial result: a polynomial distribution. 

Many years ago, Chou and Yang \cite{Chou-Yang} suggested to identify the matter distribution with that of the electric charge (form factor) in a model of elastic hadron scattering.   

We instead suggest to proceed "inversely": we identify the matter distribution with the Fourier-Bessel image of a successful elastic hadron scattering amplitude satisfying theoretical constrains and fitting the data. Matter can be scrutinized only by the interaction, the result depending also on the interaction energy, which means that our distribution will also be energy-dependent.   

The model in question combines simplicity and efficiency. It is based on a double Regge pole (here, pomeron) exchange, suggested half a centure ago \cite{Jenk}. For a recent update see Ref. \cite{Jenk3}. 

Generally, the model contains four Regge exchanges: the leading  pomeron ($P$) and odderon ($O$) \cite{Chachamis} trajectories, plus the secondary (non-leading) trajectories $f$ and $\omega$. 
Note that while $P$ and $f$ have the same $C$-parity, contributing to the $pp$ and $p \bar p$ scattering with the same sign, $O$ and
$\omega$ have different $C$-parities, with  opposite contributions to the $pp$ and $p \bar p$ scattering. We use the norm, where  
$${d\sigma_{\rm el}\over{dt}}(s,t)={\pi\over s^2}|A(s,t)|^2$$ and $$\sigma_{\rm tot}(s)={4\pi\over s}\,{\rm Im} \, A(s,t=0).$$
The invariant (spinless) elastic scattering amplitude is \cite{Jenk, Jenk3}
\begin{equation}
A\left(s,t\right)_{pp}^{p \bar p}=A_P\left(s,t\right)+A_f\left(s,t\right)
\pm\left[A_{\omega}\left(s,t\right)+A_O\left(s,t\right)\right]. \nonumber \\ \label{Eq:Amplitude}
\end{equation}

\begin{figure}[h]
\centering
		\includegraphics[width=1\textwidth]{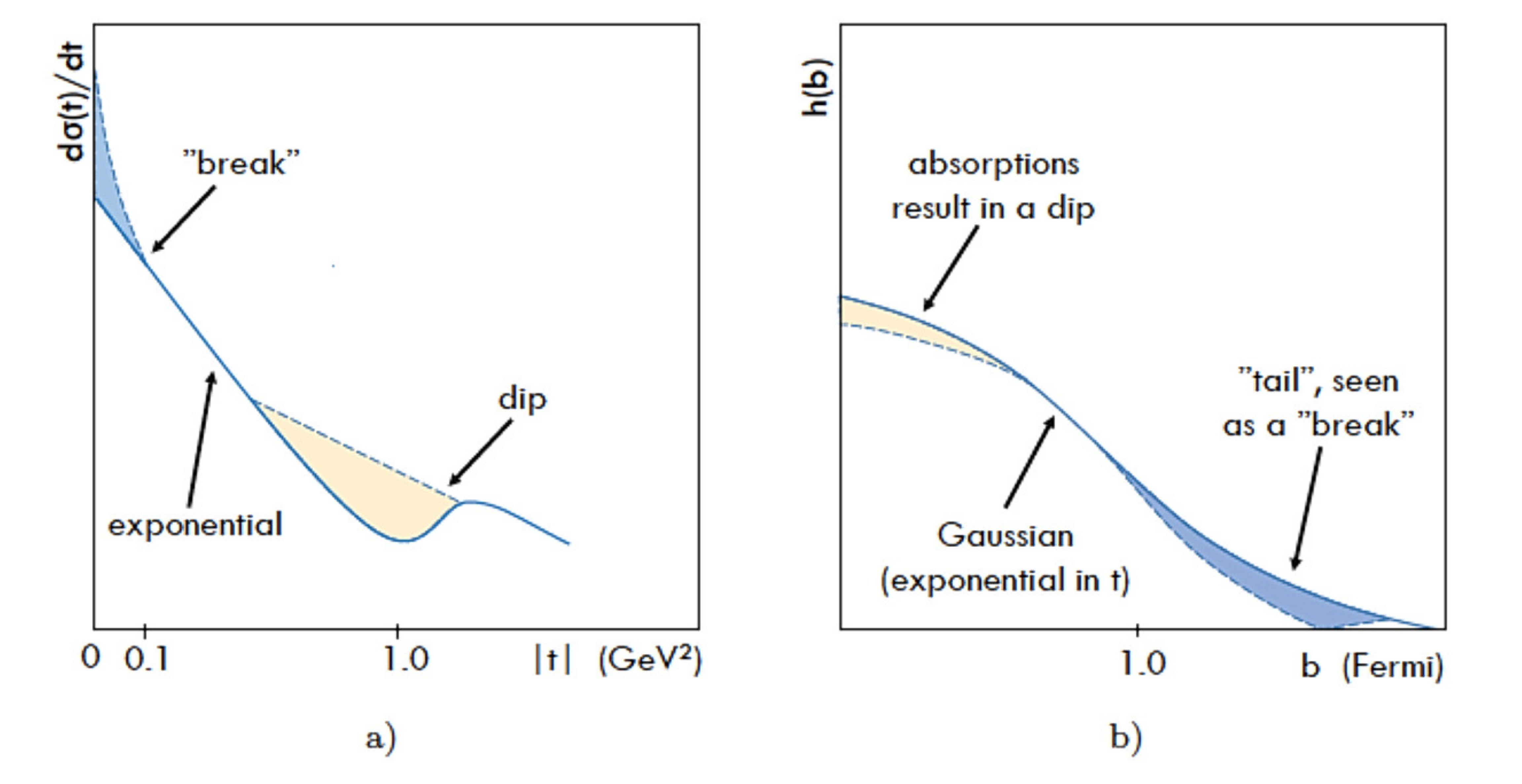}
	\caption
  {Typical pattern of high-energy proton-proton elastic differential cross-section (a) and of the corresponding proton profile, its F-B image (b). The small-$t\approx -0.1$ GeV$^2$ "break" is a consequence \cite{Cohen} of the threshold singularity in the trajectory/amplitude, as shown in the diagram, Fig. \ref{Fig:Diagram}, producing the proton atmosphere (right).} 
	\label{Fig:3}
	\end{figure}

The pomeron $A_P(s,t)$ is a dipole in the complex angular momentum plane, 
\begin{eqnarray}\label{Pomeron}
& &A_P(s,t)={d\over{d\alpha_P}}\Bigl[{\rm e}^{-i\pi\alpha_P/2}G(\alpha_P)\Bigl(\frac{s}{s_{0P}}\Bigr)^{\alpha_P}\Bigr]= \nonumber \\
& &{\rm e}^{-i\pi\alpha_P(t)/2}\Bigl(\frac{s}{s_{0P}}\Bigr)^{\alpha_P(t)}\Bigl[G'(\alpha_P)+\Bigl(L_P-{\frac{i\pi}
{2}}\Bigr)G(\alpha_P)\Bigr].
\label{Pomeron}\end{eqnarray}

We fix the first term in squared brackets since it determines the diffraction cone:
\begin{equation} \label{residue} G'(\alpha_P)=-a_P{\rm
	e}^{b_P(\alpha_P-1)},
\end{equation} 
where $G(\alpha_P)$ is recovered by integration. Moreover it is  \mbox{$L_P\equiv\ln(s/s_{0P})$}. Eq.~(\ref{Pomeron}) can be written in the following "geometrical" form \cite{Jenk}:
\begin{eqnarray}\label{GP}
A_P(s,t)&=&i{a_P\ s\over{b_P\ s_{0P}}}\Bigl[r_{1P}^2(s){\rm e}^{r^2_{1P}(s)(\alpha_P-1)}\\ \nonumber
& &-\varepsilon_P r_{2P}^2(s){\rm e}^{r^2_{2P}(s)(\alpha_P-1)}\Bigr],
\end{eqnarray} 
where $r_{1P}^2(s)=b_P+L_P-i\pi/2$, $r_{2P}^2(s)=L_P-i\pi/2$ 
and (in linear approximation)
\begin{eqnarray}\label{Ptray}
\alpha_P\equiv \alpha_P(t)=1+\delta_P+\alpha'_Pt
\end{eqnarray}
is the "supercritical"\ , {\it i.e.}
 $\alpha_P \equiv \alpha_P(t)>1$ pomeron trajectory. The fitted parameters may vary, depending on the kinematical region covered, precision of fits, form of the (non-linear) trajectories (see, for instance, Ref. \cite{Trajectory}), inclusion of non-leading terms, etc. For example, in early (70-ies) fits
 (see Ref. \cite{Jenk}), it was obtained  $\delta_P=0.017,\ \ \epsilon_P=0.012,\ \ b_P=19.42,\ \ s_{0P}=16.46$ GeV$^2$. Updates can be found e.g. in Ref. \cite{Jenk1} and references therein.

Remind that the spatial (impact parameter) distribution of the nuclear matter is obtained by a Fourier-Bessel integration:
\begin{equation}
		h(b)=\int_0^\infty d\sqrt{-t}\sqrt{-t}J_0(b\sqrt{-t})A(t),
	\end{equation}
as shown in Fig. \ref{Fig:3}. 

	\section{Generalized Parton Distributions}\label{Sec:GPD}
\label{FormF}
Generalized parton distributions (GPDs) are not measured directly, instead they can be represented as convolution integrals like the following:
\begin{equation}\label{GPD}
A(\xi, t, Q^2)=\int\limits_{-1}^1\frac{\rm{GPD}(x,\xi, t, Q^2)}{x-\xi+i\varepsilon}dx,
\end{equation}
where $\xi\approx\frac{x_{B}}{2-x_{B}}$ is called skewedness, $Q^2$ is the photon virtuality (see Fig. 1) and 
 $x$ is the fraction of the parton longitudinal momentum. The integral corresponds to the deeply virtual Compton scattering (DVCS) and $x$ is the integration variable (not to be confused with the  Bjorken variable $x_{B}$).
 The above is an integral equation for the unknown function GPD. Its solution is well known along the line $x=\xi$, where 
\begin{equation}\label{GPD1}
 \rm{GPD}(x=\xi,\xi,t,Q^2)=-\frac{1}{\pi}\rm{Im}A(\xi,t,Q^2).
\end{equation}
 
 The function GPD (nominator in Eq. (\ref{GPD})) is universal in any exclusive reaction of the type  $\gamma^*p\rightarrow Vp,$ 
 where $V$ denotes a real photon or a vector meson. Accordingly, various exclusive processes of the vector particle production can be described by means of Eq. (\ref{GPD}) with a universal function GPD.
 
Since a complete deconvolution of Eq. (\ref{GPD}) is impossible, (see {\it e.g.} Ref. \cite{Mueller_Kumericki}), one uses models and approximations. We will focus on the dependence of the GPDs on the fraction of the longitudinal momentum $x$ of the parton and the squared momentum transfer $t$, related to the matter distribution in the transverse plane (let us denote this function 
$\cal H)$. They relate the partonic distributions $q(x)$ with charge form factors $F(t)$. 
  
The Fourier-Bessel integral 
\begin{equation} \label{FB}
q(x,b)={1\over 2\pi}\int_0^\infty
d\sqrt{-t}J_0(b\sqrt{-t}){\cal H}(x,t)\end{equation} 
provides for a mixed representation of the longitudinal and transverse position of partons in the infinite momentum system. 

We assume a simple model of parton densities in nucleons in which the function $\cal H$ is factorized according to the ansatz
\begin{equation} \label{Factor} 
	{\cal H}(x,t)=q_v(x)F_1(t),
\end{equation}
where $q_v(x)$ is the (valence) DIS distribution function \cite{Watanabe} and $F_1(t)$ is the proton form factor. This model, however, contradicts both the Regge behaviour  
\begin{equation}
	{\cal H}_R(x,t)\sim x^{-\alpha(t)},
\end{equation}
and the light cone formalism implying the Gaussian ($G$) parametrization 
 \begin{equation} \label{G}
	{\cal H}_G^q(x,t)=q_v(x){\rm e}^{-(1-x)t/4x\lambda^2},
\end{equation}
where $\lambda^2$ is the typical transverse momentum of valence quarks in the nucleon. 

To satisfy the Drell-Yan-West condition
 \cite{DY, West}, that constrains the behavior of the structure functions at $x\to 1$ as well as the $t-$dependence of elastic form factors, the above expression (\ref{G})
 should be modified {\it e.g.} as in Ref. \cite{Jenk}: 
\begin{equation} \label{b}
{\cal H}_G^q(x,t)=q_v(x)x^{-\alpha'(1-x)t},
\label{alpha_tilde}\end{equation}
where $\alpha'$ is the slope of the Regge trajectory. 
 
Following Ref. \cite{Jenk} we write the $t-$dependent part ${\cal H}^q_G$ of the function GPD as 
\begin{equation} \label{alpha_tilde}
{\cal H}_G^q(x,t)=q_v(x)(x/g_0)^{-\tilde\alpha(t)(1-x)t}=(x/g_0)^{-\alpha(t)(1-x)}f(x),
\label{alpha_tilde}\end{equation}
where  $g_0>1$ is a free parameter, $\tilde\alpha(t)$ is the $t-$ dependent part of the trajectory and  
$f(x)$ is the reaction-dependent parton distribution function that we do not specify for the moment, concentrating on the interplay between the $x$ and $t$ dependences of the function GPD.

We follow the idea put forward in Ref. \cite{Jenk1} according to which the linear dependence on  $t$ in Eq. (\ref{b}) is a 
"rudiment"~of the Regge trajectory, corresponding to its linear part. Correspondingly, the exponent in Eq. (\ref{b}) denoted as  $\tilde\alpha(t)$ is $\alpha(t)-\alpha(0)$, being the expression of    
$\alpha(t)$ defined by Eq.~(\ref{3}), quoted in Sec.  2.

\begin{figure}[h] 
	\centering
	\includegraphics[width=1\textwidth]{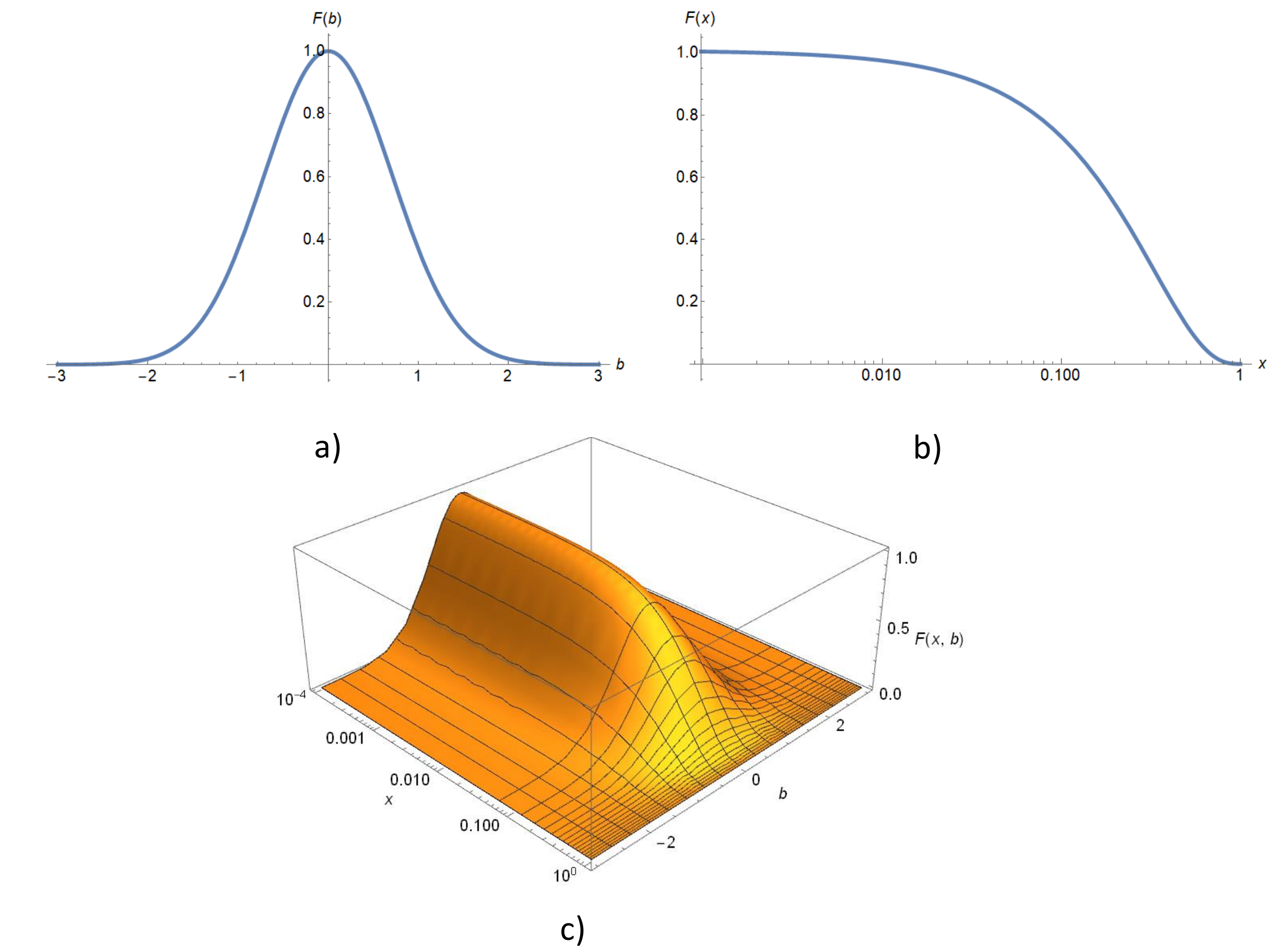}
	\caption{Distribution of matter in the nucleon a) in the impact parameter $b$ (Gaussian in the expression (\ref{Gauss})), b) in the Bjorken variable $x$, c) in both dimensions on the "tomographic" 3D picture. Only the gluon component (\ref{f}), which dominates at small values of $x$, was taken into account.} 
	\label{Fig:3d}
\end{figure}

Fig. \ref{Fig:3d} shows the resulting transverse (a) and longitudinal (b) parts of the parton distribution function as  well as both in the three-dimensional picture (c).

We plan to give detailed calculations in a next work, and for now we give a qualitative three-dimensional picture for the distribution of gluons in the proton in a simplified factorized model.

For the gluon distribution we assume the Gaussian form
\begin{equation}\label{Gauss}
	F(x,b)=f_P(x){\rm e}^{-b^2/R^2},
\end{equation}
which is a Fourier image of Eq. (\ref{Factor}), corresponding to the model presented in the Sec. \ref {Sec:Matter}. This will be explored in our next paper. In turn, for the distribution function of gluons we use the simple expression 
\begin{equation}\label{f}
	f_P(x)=x^{-\alpha(0)}(1-x)^N,
	\end{equation}
where 
$N$ is the number of partons (gluons or quarks) in the nucleon. For us, this number (for example,$г\approx 2$ for gluons or $\approx 3$ for quarks) is not significant in comparison with $ 1+\delta $, which determines the behavior of parton distributions at small $x$, where there is the problem of saturation of parton density and where possible reverse process of parton recombination is known. In this model, we obtained the distribution of matter in the proton, shown in Fig. \ref {Fig:3d}, where the right extreme three-dimensional picture is important without detailing the behavior at small and large $x$.

\section{Pressure distribution}\label{Sec:Pressure}

The mechanical properties of the proton are encoded in TEI. Direct determination of the distribution of the quark pressure in the proton requires the calculation of the element of the proton matrix of the energy-momentum tensor \cite{Polyak1, Polyak2, Teryaev}.  One of the GFFs, $d(t)$,  contains information about the distribution of forces and pressure of 
quarks in the proton within the generalized parton distributions (GPDs), which can be accessed experimentally in the DVCS. 

\begin{figure}\label{CLAS}
		\centering{\includegraphics[width=0.6\textwidth]{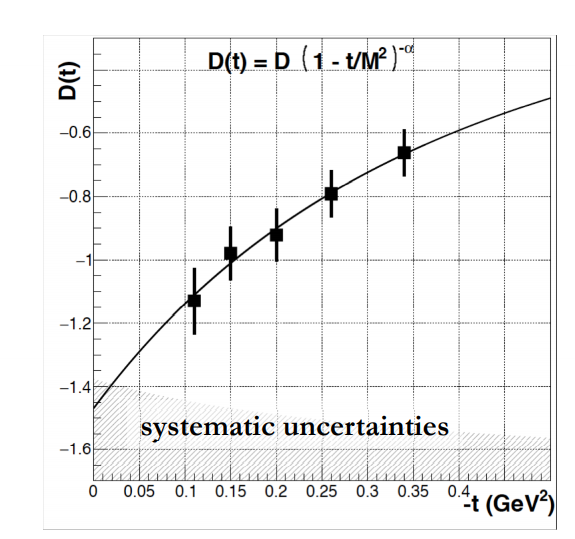}}
		\caption{$D(t)$ from CLAS 6 GeV data \cite{Burkert_Elouadrhiri_Girod}} \vspace{0.5cm}
		\label{Fig:Diagram111}
	\end{figure}

	In Ref. \cite{Burkert_Elouadrhiri_Girod} $ d (t) $ was obtained from data of the polarized and unpolarized DVCS scattering $ep$, measured by the CLAS collaboration at 
	JLab, at the electron beam energy of $6$ GeV \cite{CLAS1, CLAS2}.  The following empirical parametrization was used for comparison with the data:
	\begin{equation} \label{alpha}
		d(t)=d(0)\Bigl(1-{\frac{t}{M^2}}\Bigr)^{-\alpha},
	\end{equation}
	where the free parameters $d(0)$ and $\alpha$ were adjusted to the data, and the index 
	$\alpha=3$  was fixed from the quark counting rules; for instance, Fig. \ref{CLAS} shows the graph of the so called   
	"subtraction term" D(t), proportional to the d(t) (see below the relation between the two quantities). M is another adjustable parameter.  Take into account that the parameter $\alpha$ can be associated with the asymptotic (logarithmic) behavior of the leading Regge trajectory \cite{Jenk}.

Further measurements of this parameter at the electron beam energy of $12$ GeV are 
planned at JLab (see, for instance, Ref. \cite{CLAS3}).

\begin{figure}[tbh!]
\centering{\includegraphics[width=0.8\textwidth]{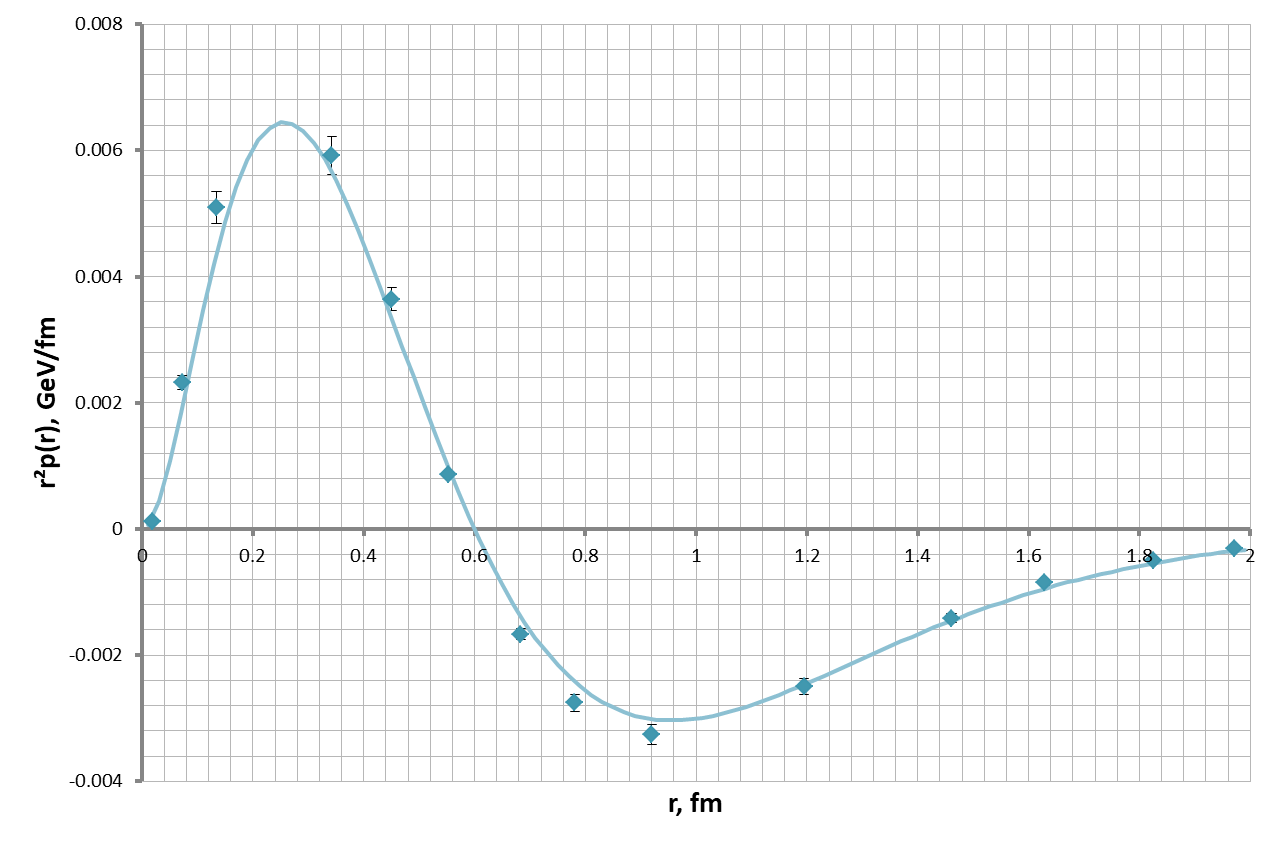}}
\caption{The curve $r^2p(r) $, where  $p(r)$ is the pressure distribution inside the proton according to the formula (\ref{4}). It is compatible with the experimental data \cite{Burkert_Elouadrhiri_Girod}.
} \vspace{0.5cm}
\label{Fig:pressure}
\end{figure}

The pressure $p(r)$ can be determined from the Fourier-Bessel integral
\begin{equation}
	p(r)=\int_0^\infty d\sqrt{-t}\sqrt{-t}J_0(r\sqrt{-t})d(t).
\end{equation} 

Parametrizations of $r^2p(r),$ associated with quark bag models (see, {\it e.g.},  Ref. \cite{Jenk}) are of interest. The pressure must satisfy the von Laue stability condition \cite{Laue}:
\begin{equation} \label{Laue}
	\int_0^\infty r^2p(r)dr=0.
\end{equation}
We have studied empirical models of $p(r)$ satisfying this condition.

A simple parametrization, namely the sum of theta and delta functions, was studied by Polyakov and co-authors \cite{Polyak1,Polyak2}. It reads
\begin{equation} \label{bag}
	p(r)=p_0\Theta(R_d-r)-p_0R_d\delta(R_d-r),
\end{equation} 
and corresponds to a proton with (unrealistic) sharp edges.

We instead suggest the following empirical formula for the pressure:
\begin{equation}\label{4}
	p(r)=A(B-r){\rm e}^{-{r/a}}, 
\end{equation}
where $A=1$ GeV fm$^{-3}$ is a coefficient, $B=0.6$ fm is a point taken from experimental data at which $p(r)$ = 0, the value $a$ = 0.2 fm was chosen such as to provide that the maximum and minimum in Fig. \ref {Fig:pressure} meet the data.

According to Ref. \cite{Burkert_Elouadrhiri_Girod} and references therein, the so called "subtraction term" $D(t)$ and the GFF form factor $d(t)$ are related as 
\begin{equation} \label{dt}
	d(t)=\frac{25}{18}D(t).
\end{equation}
Furhtermore, having fixed $D(0)=-1.5$ (see Fig. \ref{Fig:Diagram111}  in Ref. \cite{Burkert_Elouadrhiri_Girod}), one gets $d(0)\simeq-2$. We remind that  $p(r)$ and $d(t)$ are related by
\begin{equation} \label{pr}
	d(t)=\int\frac{j_0(r\sqrt{-t})}{2t}p(r)\,\mathrm{d^3}r.
\end{equation}
By substituting the expression  (\ref{4}) of $p(r)$ into Eq.  (\ref{pr}) we obtain
 \begin{equation} \label{dt1}
 	d(t)=\frac{-2}{(1-{t\over{25}})^3}.
 \end{equation}

Alternatively, from Eq.~(\ref{alpha}) with $\alpha = 3$, $d(0) = -2$ and $M^2 = 25$ we get a similar behaviour, confirming that  formula (\ref {4}) is a reasonable approximation to the experimental data.
 
To conclude, our parametrization fits the data and satisfies the von Laue stability condition. 

\section{Conclusions}
Interest in the study of gravitational form factors is connected both with theoretical works \cite{Polyak1, Polyak2} and with recent measurements of DVCS processes on the JLab electron accelerator. Recall that this is not a direct gravitational interaction with nucleon quarks, which, however, can be studied in hard processes of quantum chromodynamics (QCD). At present, preliminary results have been obtained in this direction, which allow only indirect comparisons of experimental data with explicit expressions for the pressure distribution in nucleons.

We have suggested a Reggeized model of Generalized parton distributions (GPDs) together with a realistic complex Regge trajectory, which obtained a three-dimensional picture of the distribution of nuclear matter in the proton, as well as the pressure distribution in it, taking into account its positive and negative components. The latter is a manifestation  of confining forces in the nucleon. 

Work on extension of the DVCS formalism to
incorporate analyticity (e.g. complex Regge trajectories) is in progress.

In the future, we hope to investigate alternative models of pressure distribution, fitting the data satisfying the von Laue stability condition and compatible with the GPDs, using non-linear Regge trajectories.

\section*{Acknowledgements}
We thank Oleg Selyugin, Oleg Teryaev and Alessandro Papa for useful discussions as well as Alexander Korchin for his critical remarks.  
 
L.J. was supported by grant 1030 (№ 0121U109612) "Particle and collective excitation dynamics in high-energy-physics etc.".


\begin{thebibliography}{99}
		
	\bibitem{Mueller} {\it M\"uller D. et al.} Wave functions, evolution equations and evolution kernels from light ray operators of QCD // Fortsch. Phys. 1994. V. 42. P. 101-141.
	
	\bibitem{Rad} {\it Radyushkin A. V.} Nonforward Parton Distributions // Phys. Rev. D. 1997. V. 56. P. 5524-5557. 
	
	\bibitem{Ji} {\it Ji X. D.} Off-Forward Parton Distributions // J.Phys. G. 1988. V. 24. P. 1181-1205.
	
	\bibitem{Burkert_Elouadrhiri_Girod} {\it Burkert V. D., Elouadrhiri L., Girod F. X.} The pressure distribution inside the proton, Nature. 2018. V. 557. P. 396–399.
	
	\bibitem{Jenk1} {\it Jenkovszky L. L.} Duality, analyticity and t dependence of generalized parton distributions // Phys. Rev. D. 2006. V. 74. P. 114026. 
	
	\bibitem{Jenk2} {\it Fiore R., Jenkovszky L. L., Magas V. K.} Generalized parton distributions, analyticity and crossing // Nuclear Phys. B Proc. Suppl. 2005. V. 146. P. 146-150.
	
	\bibitem {Teryaev2} {\it Teryaev O.V.} Spin structure of nucleon and equivalence principle, 1999, arXiv:9904376[hep-ph].	
	
		\bibitem {Pagels} {\it Pagels H.} Energy-Momentum Structure Form Factors of Particles // Phys. Rev. 1996. V. 144, No. 4. P. 1250-1260.
		
		\bibitem {Weinberg} {\it Weinberg S.} Gravitation and Cosmology: Principles and Applications of the General Theory of Relativity, John Wiley and Sons, 1972.
	
		\bibitem{Landau} {\it Landau M. D., Lifshits M. E.} Theory of elasticity. 4th Ed. V. 7. Rev. -M .: Science. Ch. ed. phys.-mat. lit. 1987. P. 248.
	
	\bibitem{Chugreev} {\it Chugreev Y. V.} The energy-momentum tensor in the relativistic theory of gravitation // PEPAN Letters. 2018. V. 8. No. 6(218). P. 467-475.
	
	\bibitem{Trajectory} {\it Szanyi I. et al.} Pomeron/glueball and odderon/oddball trajectories // Nuclear Physics  A. 2020. V. 998. 121728.

	\bibitem{Vorobyev} {\it Vorobyev A. A.} Precision measurements of the proton charge in electron-proton scattering // PEPAN Letters. 2019. V. 16. No. 5(224). P. 390-391.
	
	\bibitem{Selugin} {\it Selyugin O. V.} Electromagnetic and gravitomagnetic structure and radii of nucleons // EPJ Web of Conferences EPJ Web of Conferences. 2019. V. 222. 03018.
			
	\bibitem{Chou-Yang} {\it Chou T. T., Yang C. N.} Model of Elastic High-Energy Scattering // Phys. Rev. 1968. V. 170. No. 5. P. 1591.
	
		\bibitem{Jenk} {\it Wall A.N., Jenkovszky  L.L., Struminsky B.V.} High energy hadron interactions // Physics of elementary particles and the atomic nucleus. 1988. V. 19,  No. 1. P. 180-223.
	
	\bibitem{Jenk3} {\it Jenkovszky L. L., Schicker R., Szanyi I.} Elastic and diffractive scattering in the LHC era // IJMPE. 2018. V. 27, No. 08. 1830005.		
	
\bibitem{Chachamis} {\it Chachamis G., Vera A.S.} Reggeon webs, spin chains and the odderon, 2018,	arXiv:1810.06690 [hep-ph].

\bibitem{Cohen} {\it Cohen-Tannoudji G., Ilyin V. V., Jenkovsky L. L.} Model for the pomeron trajectory // Lett. Nuovo Cim. 1972. V. 5. No. 14. P. 957-962.

	\bibitem{Mueller_Kumericki} {\it Kumerički K., M\"uller D.} Description and interpretation of DVCS measurements, 2016, arXiv:1512.09014 [hep-ph].
	
	\bibitem{Watanabe} {\it Watanabe A.} DIS at small $x$ and hadron-hadron scattering at high energies via the holographic pomeron exchange, 2018, 	arXiv:1810.07474[hep-ph].
	
		\bibitem{DY} {\it ~Drell S. D., ~Yan T. M.} Connection of Elastic Electromagnetic Nucleon Form Factors at Large $Q^2$ and Deep Inelastic Structure Functions Near Threshold // Phys. Rev. Lett. 1979. V. 24.  P. 181.
		
		\bibitem{West} {\it ~West G.B.} Phenomenological Model vor the Electromagnetic Structure of the Proton // Phys. Rev. Lett. 1970. V. 24. P. 1206.	
		
		\bibitem{Polyak1} {\it Polyakov M.V.}  Generalized parton distributions and strong forces inside nucleons and nuclei // Phys. Lett. B. 2003. V. 555. P. 57-62.
		
		\bibitem {Polyak2} {\it Polyakov M. V., Schweitzer P.} Forces inside hadrons: pressure, surface tension, mechanical radius, and all that // Int. J. Mod. Phys. A. 2018. V. 33. 1830025.
		
		\bibitem {Teryaev} {\it Teryaev O. V.} Spin structure of nucleon in QCD: inclusive and exclusive processes // "New Trends in High-Energy Physics", Proc. of the 2001  Crimean conference, edited by P.N. Bogolubov et al., Kiev, 2001, p. 256. 
				
		\bibitem{CLAS1} {\it Girod F.X. et al} [CLAS collaboration] Deeply Virtual Compton Scattering Beam-Spin Asymmetries // Phys. Rev. Lett. 2008. V. 100. 162002.
		
		\bibitem{CLAS2} {\it Girod F.X. et al} [CLAS Collaboration] Cross sections for the exclusive photon electroproduction on the proton and Generalized Parton Distributions //  Phys. Rev. Lett. 2015. V. 115. 212003.		
				
		\bibitem{CLAS3} {\it Girod F.X. et al} [CLAS Collaboration] Deeply Virtual Compton Scattering with CLAS12 at 6.6GeV and 8.8 GeV. Proposal E12-16-010B. Jefferson Lab PAC44, 2016.
				
		\bibitem{Laue} {\it von Laue M.} Annalen der Physik. 1911. V. 35. N. 8, 524. P. 541. Formula (21).
	
%
%
%

\end{thebibliography}
\end{document}